\documentclass[aps,superscriptaddress,10pt,floatfix,amsmath,amssymb,showkeys]{revtex4-2}
\usepackage{graphicx} 
\usepackage{dcolumn}
\usepackage{bm}
\usepackage{physics}
\usepackage{lipsum}  
\graphicspath{{images/}}

\begin{document}

\title{On the use of superthermal light for imaging applications}
\date{\today}

\author{S. Cassina}
\email{s.cassina@uninsubria.it}
\affiliation{Department of Science and High Technology, University of Insubria, Via Valleggio 11, I-22100 Como, Italy}
 
\author{G. Cenedese}
\affiliation{Department of Science and High Technology, University of Insubria, Via Valleggio 11, I-22100 Como, Italy}
\affiliation{INFN-Section of Milan, Via Celoria 16, I-20133 Milano, Italy;}

\author{M. Lamperti}
\affiliation{Department of Science and High Technology, University of Insubria, Via Valleggio 11, I-22100 Como, Italy}
\affiliation{Institute for Photonics and Nanotechnologies, IFN-CNR, Via Valleggio 11, I-22100 Como, Italy}

\author{M. Bondani}
\affiliation{Institute for Photonics and Nanotechnologies, IFN-CNR, Via Valleggio 11, I-22100 Como, Italy}

\author{A. Allevi}
\affiliation{Department of Science and High Technology, University of Insubria, Via Valleggio 11, I-22100 Como, Italy}
\affiliation{Institute for Photonics and Nanotechnologies, IFN-CNR, Via Valleggio 11, I-22100 Como, Italy}


\begin{abstract}
Ghost imaging and differential ghost imaging are well-known imaging techniques based on the use of both classical and quantum correlated states of light. Since the existence of correlations has been shown to be the main resource to implement ghost imaging and differential ghost-imaging protocols, here we analyze the advantages and disadvantages of using two different kinds of superthermal states of light, which are more correlated than the typically employed thermal states. To make a fair comparison, we calculate the contrast (C) and the signal-to-noise ratio (SNR) of the reconstruct image. While the larger values of C suggest the usefulness of these superthermal states, the values of SNR do not improve by increasing the intensity fluctuations of light. On the contrary, they are the same as those exhibited by thermal light.
\end{abstract}
\maketitle

Imaging techniques, such as ghost imaging (GI) and differential ghost imaging (DGI), are based on the use of bipartite correlated states of light: one arm illuminates the object, while the other arm reconstructs the image \cite{dangelo,gatti,ferri05,ferri10}. Over the years, it has been demonstrated that the main resource to obtain such ghost images is not given by the entangled nature of the light source \cite{pittman,abouraddy01,abouraddy04}, but rather on the presence of intensity correlations \cite{gatti04,gatti04a,bache04,bache04a,crosby,arimondo}. Indeed, both classical and quantum light sources can be exploited, even if some differences exist, usually quantified in terms of contrast (C) and signal-to-noise ratio (SNR) \cite{shapiro}. It is possible to show, both theoretically and experimentally, that the highest values of C and SNR can be achieved by using quantum-correlated states of light, even if they are more fragile and sensitive to loss \cite{brida,epj12,losero}.\\
Another critical issue is represented by the strength of correlations \cite{gatti06}: if the main resource is the existence of correlations, it is crucial to understand whether higher correlations mean higher values of C and SNR, and thus higher-quality images \cite{shapiro,samantaray}. To address the problem, in this Letter we compare the case of thermal light, that is the standard light source employed in ghost-imaging protocols \cite{ferri05,iskhakov}, with two different kinds of optical states endowed with higher-than-thermal intensity fluctuations, and thus called superthermal light states \cite{bianciardi,OL15}. While thermal states can be generated by passing a laser beam through a rotating ground-glass disk \cite{arecchi,OLcorr}, superthermal ones can be produced applying either linear manipulations \cite{odonnell,newman,gori} or nonlinear interactions \cite{qu} to a thermal field. 

\begin{figure}[hbtp]
    \centering
    \includegraphics[scale =0.4]{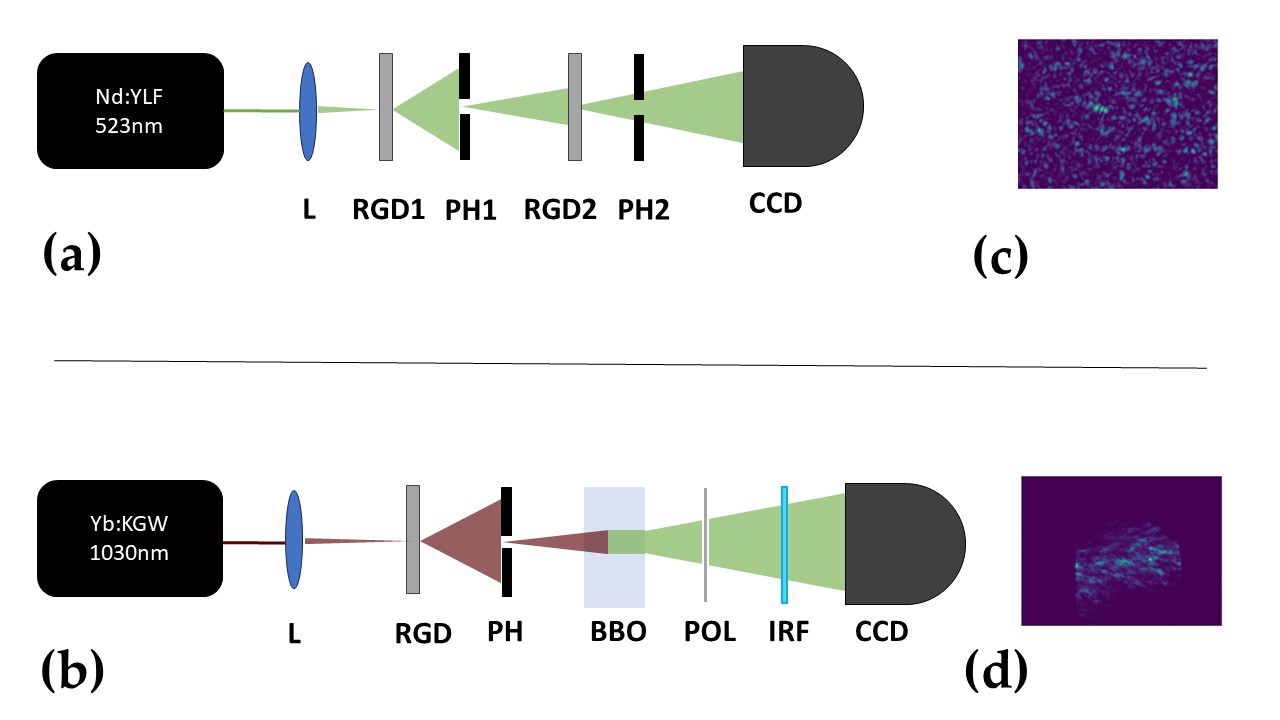}
    \caption{(a) Sketch of the experimental setup in case A; (b) Sketch of the experimental setup in case B. L: convergent lens; RGD1, RGD2, RGD: rotating ground-glass disk; PH1, PH2, PH: pin-hole; CCD: charge coupled device; POL: polarizer; IRF: infrared filter. (c) and (d): typical single-shot image in case A and B, respectively. See the text for details.}
    \label{fig:setup_PLA}
\end{figure}

\begin{figure}[hbtp]
    \centering
     \includegraphics[scale =0.55]{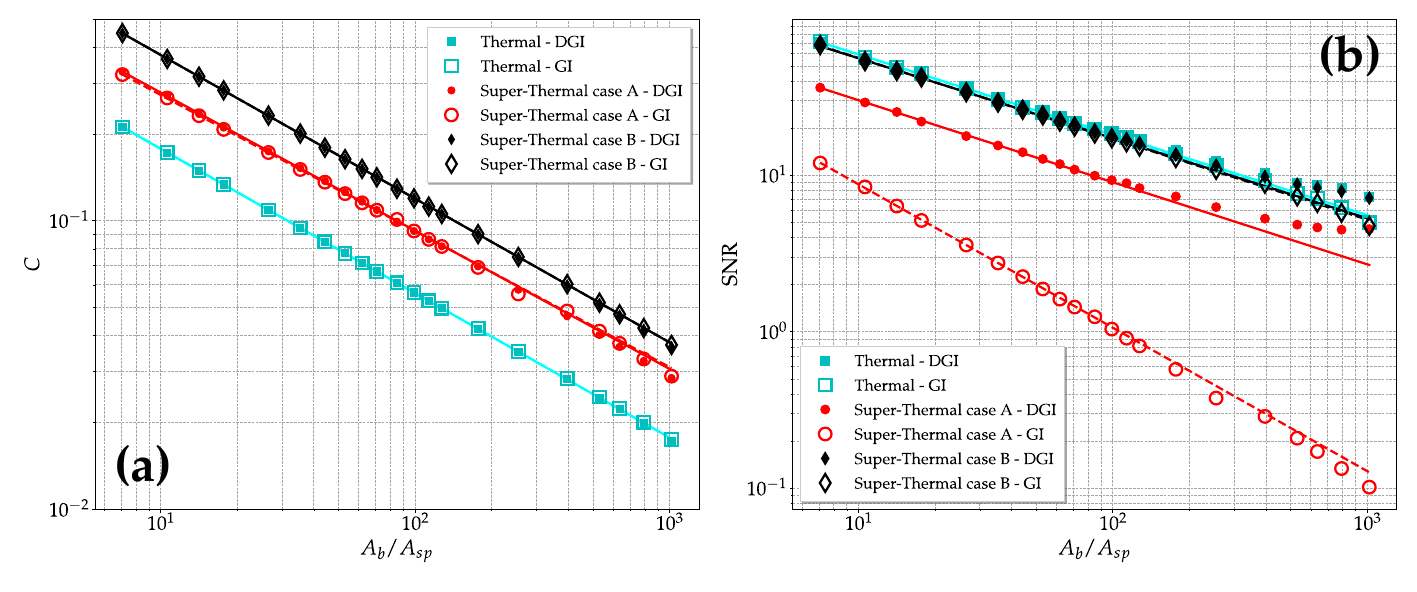}
    \caption{(a) C and (b) SNR obtained from simulated images as functions of the ratio between the area of the object, $A_b$, and that of a typical speckle, $A_{sp}$, which roughly corresponds to the number of speckles illuminating the object. Cyan symbols: results  corresponding to thermal light; red symbols: results corresponding to case A; black symbols: results corresponding to case B. Full symbols refer to DGI, while open symbols to GI. Colored lines: fitting
functions $y = ax^b$, with $a$ and $b$ as free-fitting parameters. Solid lines correspond to full symbols, while dashed lines to open
ones. The color choice is the same as symbols. In all cases, except for GI in case A, the curves are well fitted by a power law with exponent $b\simeq 0.5$.}
    \label{fig:confrontoSNRandC}
\end{figure}

\begin{figure}[h!]
    \centering
     \includegraphics[scale =0.55]{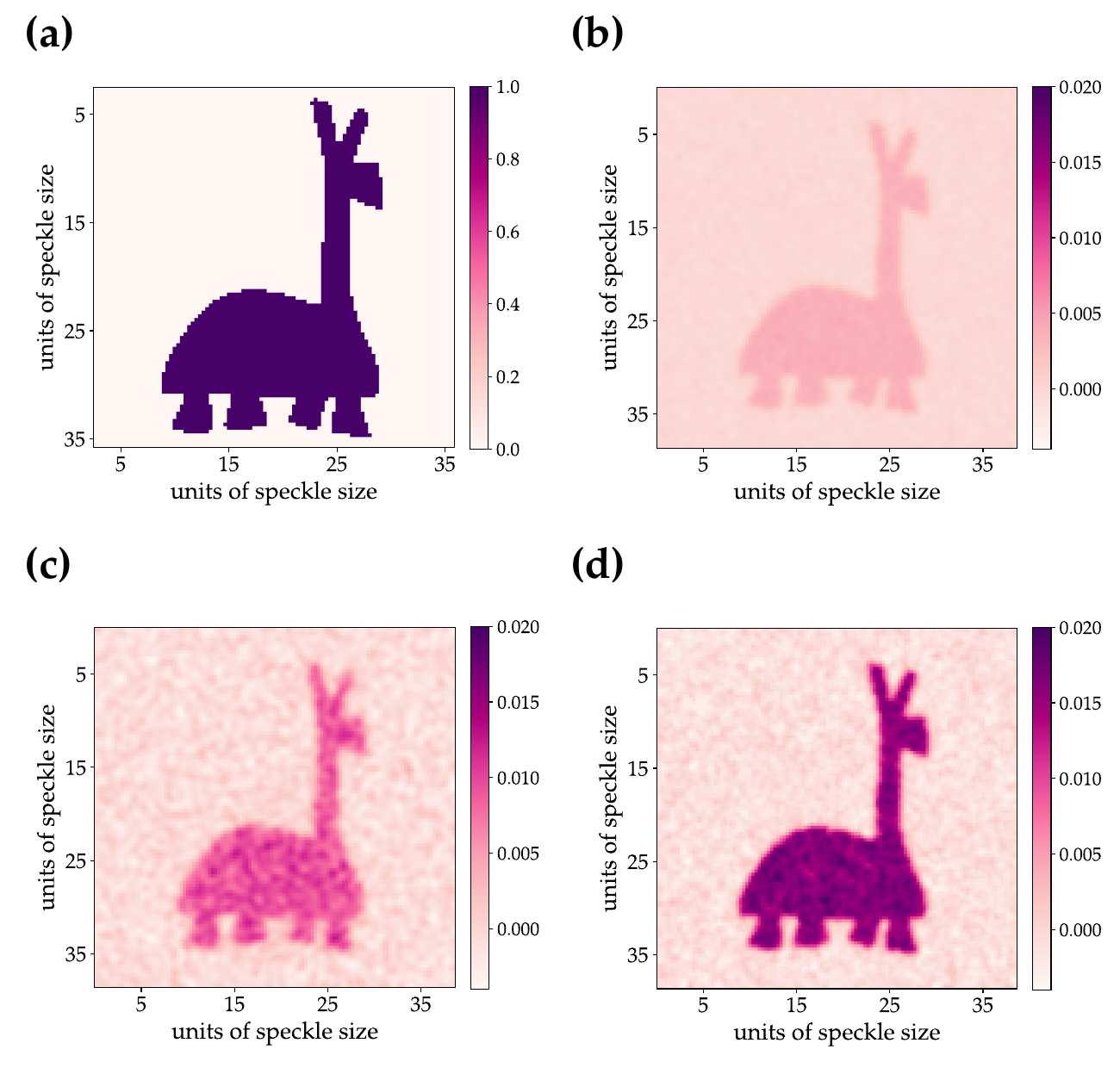}
    \caption{(a) Binary mask of a llama; (b)-(d): DGI images obtained from thermal light, superthermal light in case A and superthermal light in case B, respectively, shown with the same dynamical range.}
    \label{fig:DGIsimulfixed}
\end{figure}

\begin{figure}[h!]
    \centering
     \includegraphics[scale =0.55]{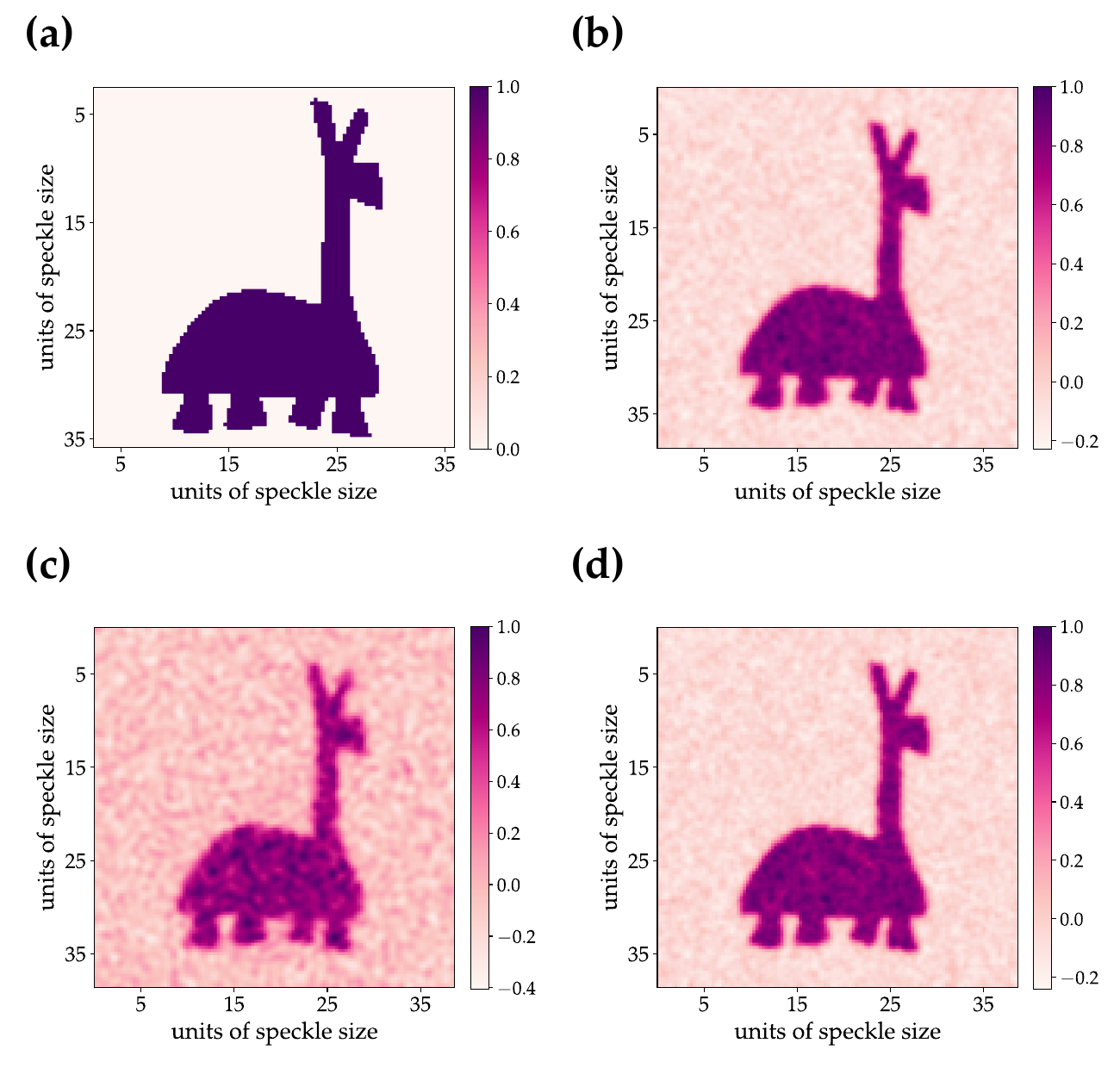}
    \caption{(a) Binary mask of a llama; (b)-(d): DGI images obtained from thermal light, superthermal light in case A and superthermal light in case B, respectively. Each DGI image is shown normalized to its maximum.}
    \label{fig:DGIsimulnorm}
\end{figure}

The intensity probability distributions corresponding to the two kinds of superthermal light sources can be obtained starting from the intensity distribution of the thermal field, which is given by \cite{mandel}
\begin{equation} \label{Pthermal}
    P_{\rm th}(I) = \frac{1}{\langle I \rangle} \exp\left({\frac{-I}{\langle I \rangle}}\right),
\end{equation}
where $\langle I \rangle$ is the mean value of the distribution. This field is composed of speckles, namely coherence areas that are the result of the interference of the radiation coming from the scattering centers of the disk illuminated by the laser light \cite{goodman}. 
In the first case of superthermal source, called case A hereafter, $\mu_f$ speckles of this thermal field are selected by a pin-hole and sent to a second rotating ground-glass disk, thus producing a speckled-speckle field. If $\mu_s$ speckles of this field are then selected by a second pin-hole, the light intensity is characterized by the following superthermal statistics \cite{goodman}
\begin{equation}\label{densityfunctionsuper}
    P_{A}(I) = \frac{2(\mu_f \mu_s)^{\frac{(\mu_f + \mu_s)}{2}}}{\langle I \rangle \Gamma(\mu_s) \Gamma(\mu_f)} \left( \frac{I}{\langle I \rangle} \right)^{\frac{(\mu_f+\mu_2-2)}{2}} K_{|\mu_f-\mu_s| }\left( 2\sqrt{\mu_s \mu_f \frac{I}{\langle I \rangle} } \right),
\end{equation}
where $\langle I \rangle$ is the mean value of the distribution and $K_{|\mu_f-\mu_s|}$ is the $|\mu_f-\mu_s|$-order modified Bessel function of the second kind.\\
From this statistics it is possible to calculate all the moments of the distribution. For imaging applications, it is essential to consider the second-order autocorrelation function, which results \cite{bianciardi}
\begin{equation}\label{g2twodisks}
    g^2(I) = \frac{\langle I^2 \rangle}{\langle I \rangle^2} = \left( 1+ \frac{1}{\mu_f} \right)\left( 1+ \frac{1}{\mu_s} \right).
\end{equation}
The maximum value of the function, $g^2_{\rm MAX}(I) = 4$, is expected for $\mu_f = \mu_s = 1$. We also notice that for the same choice of $\mu_f$, the minimum value is equal to $g^2_{\rm MIN}(I) = 2$, attained for $\mu _s \rightarrow \infty$ (or $\mu _s \gg 1$ ). We have already demonstrated that this value constitutes the background of the ghost image \cite{cassina23}. This means that both the maximum and minimum values of $g^2$ differ from the case of thermal light, for which $g^2_{\rm MAX} = 2$ and $g^2_{\rm MIN} = 1$.\\
The situation is slightly different in the case of the other superthermal light we consider, namely the one generated by exploiting nonlinear interactions. In this case, called case B hereafter, $\mu$ speckles of the speckle field generated by the rotating ground-glass disk are selected by a pin-hole and frequency doubled in a second-order nonlinear crystal. The resulting intensity distribution is described by \cite{OL15}
\begin{equation}\label{PsuperSH}
    P_{\rm B}(I) = \frac{b^{(\mu-2)} \exp{\left[-\mu b/\langle I_{\rm F} \rangle\right]}}{2k(\mu-1)!(\langle I_{\rm F} \rangle/\mu)^{\mu}},
\end{equation}
where $\langle I_{\rm F} \rangle$ is the mean value of the fundamental intensity distribution and $b = \sqrt{I/k}$, in which $k$ is the conversion efficiency.
The corresponding second-order autocorrelation function reads as \cite{QMQM}
\begin{equation}\label{g2SH}
    g^2(I) = 1+ \frac{2 (2 \mu +3)}{\mu (\mu + 1)}.
\end{equation}
Hence, the maximum value of the function is $g^2_{\rm MAX} = 6$, while the minimum, corresponding to the case in which $\mu \rightarrow \infty$, is  $g^2_{\rm MIN} = 1$. This means that the maximum value of $g^2(I)$ reachable with superthermal light from second harmonics is larger than the value obtained from two disks, while the minimum is the same as thermal light.
In order to experimentally verify all these features, we realized two different experimental schemes: in panel (a) of Fig.~\ref{fig:setup_PLA} we show the setup corresponding to case A, while in panel (b) of the same figure we plot the setup corresponding to case B. In Fig.~\ref{fig:setup_PLA}(a)
the second-harmonic pulses (7 ps pulse duration with a repetition rate of 500 Hz) of a Nd:YLF laser at 523 nm impinge on a rotating ground-glass disk (RGD1) and generate a pseudo-thermal field. A pin hole (PH1) allows the selection of a single speckle of this field. The selected light is scattered by a second rotating ground-glass disk (RGD2), completely uncorrelated to the first one. The output field is then selected by a second pin-hole (PH2) and measured by means of a CCD camera (DCU223M, Thorlabs, 1024 $\times$  768 squared pixels, 4.65 $\mu$m pixel pitch). A typical single-shot image of the speckled-speckle field is shown in panel (c) of Fig.~\ref{fig:setup_PLA}.
In panel (b) of the same figure, we show the setup used for the generation of super-thermal light obtained as the second harmonics of a thermal field. More precisely, the pulses (190 fs pulse duration with a repetition rate of 500 Hz) of a Yb:KGW laser at 1030 nm are scattered by a rotating ground-glass disk (RGD) and a portion of the speckle field is selected by a pin-hole (PH) and frequency-doubled in a $\beta$-barium borate crystal (BBO), while a polarizer is inserted beyond the crystal to select second-harmonic light and remove possible spurious light at the same wavelength. Finally, in front of the CCD, a band-pass filter is inserted to avoid the residual infrared light. A typical single-shot image of superthermal light from second harmonics is shown in panel (d) of Fig.~\ref{fig:setup_PLA}.\\
In both cases A and B, we recorded $10^5$ single-shot images.\\
In order to characterize the statistical properties of the generated light states, we evaluate the averaged autocorrelation image exploiting the Fast Fourier Transform \cite{cassina23}.
\begin{figure}[hbtp]
    \centering
     \includegraphics[scale =0.6]{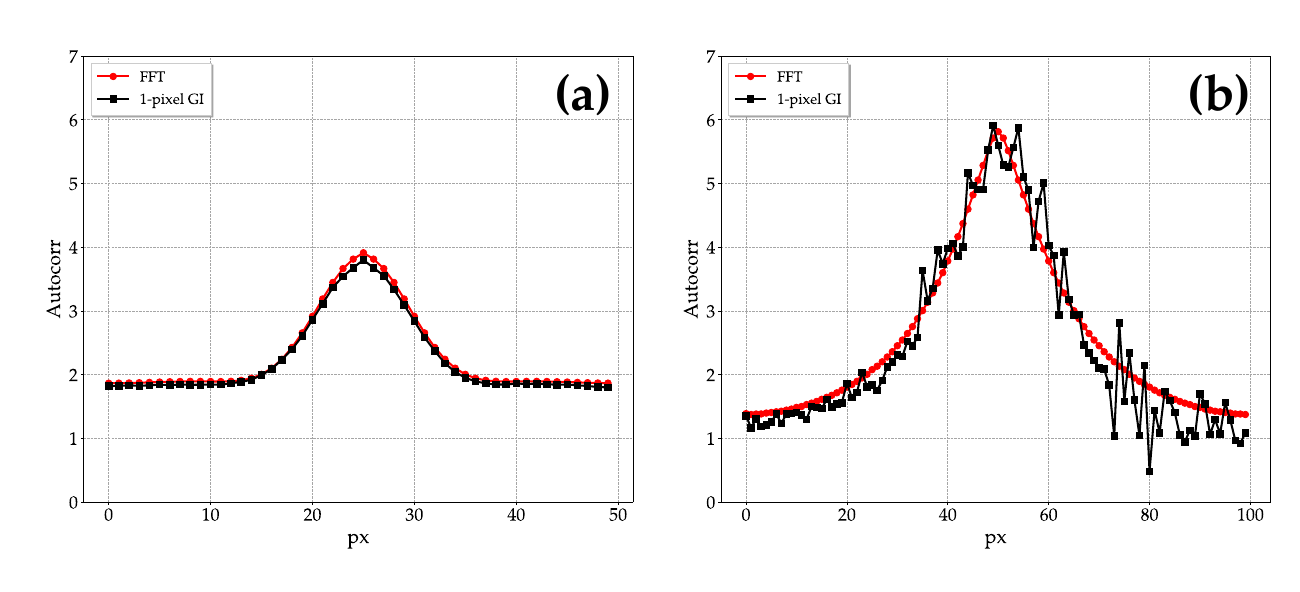}
    \caption{(a) Section of the autocorrelation function in case A; (b) the same in case B. Red curves correspond to the autocorrelation image obtained exploiting the Fast Fourier Transform, while the black dots to the autocorrelation matrix defined in Eq.~(\ref{Gpixel}).}
    \label{fig:confrontoEXPsuper}
\end{figure}
In Fig.~\ref{fig:confrontoEXPsuper}(a) we show, as red curve, a section of the autocorrelation in case A, while in panel (b) of the same figure a section of it in case B. In the same figure we present, as black dots in both panels, a section of the autocorrelation matrix obtained by correlating a single pixels with all the pixels of the CCD sensor. This is defined as
\begin{equation}\label{Gpixel}
     {\bf G_{px}}= \frac{\langle I_{px}\bf{I_{field}}\rangle}{\langle I_{px}\rangle \langle \bf{I_{field}}\rangle},
\end{equation}
 in which $I_{px}$ is the intensity corresponding to a single pixel, while $\bf{I_{field}}$ is the matrix corresponding to a single-shot image, and the averaging is obtained over $10^5$ consecutive images.\\
First of all, we notice that the sections of the autocorrelation functions calculated by means of the two methods are in very good agreement. Secondly, by comparing the curves in case A with those in case B, we can easily appreciate that the maximum and minimum values correspond to the theoretical expectations discussed above. In particular, as already anticipated, the maximum  and minimum values of $g^2$ are attained in case B. In fact superthermal light from two disks has a correlated background, and thus its minimum value of $g^2$ differ from that of case B and thermal light. Finally, we note that in panel (b) the larger fluctuations of the data are due to the  lower intensity level of the generated light in case B with respect to that in case A.\\
The results in Eq.~(\ref{Gpixel}) can be generalized to the case in which, instead of considering a single pixel, we deal with an object that occupies a certain number of pixels. This situation corresponds to the autocorrelation image in a standard GI scheme.  In this case, the GI autocorrelation matrix reads as 
\begin{equation}\label{GImask}
        {\bf{G_{GI}}}= \frac{\langle I_{bucket}\bf{I_{field}}\rangle}{\langle I_{bucket}\rangle \langle \bf{I_{field}}\rangle},
    \end{equation}
where $I_{bucket}$ is the sum of the intensities of the pixels illuminated by the light coming from the object. 
Some years ago, Ferri et al. \cite{ferri10} demonstrated that the same scheme can be used to perform the so-called DGI technique. In their work they proved that the implementation of DGI instead of GI allows for better values of the figures of merit that characterize the good quality of images. In particular, it has been verified, both theoretically and experimentally, that larger values of SNR can be obtained in the case of weakly absorbing objects. By analogy with Eq.~(\ref{GImask}), the DGI autocorrelation matrix can be written as 
    \begin{equation}\label{DGI}
         {\bf{G_{DGI}}}= \frac{\langle I_{bucket}\bf{I_{field}}\rangle}{\langle I_{bucket}\rangle \langle \bf{I_{field}}\rangle} - \frac{\langle I_{field}\bf{I_{field}}\rangle}{\langle I_{field}\rangle \langle \bf{I_{field}}\rangle},
    \end{equation}
where $I_{field}$ is the sum of the intensities detected in a portion not including the object.\\
In some previous papers of ours \cite{cassina23,QMQM} it was demonstrated that superthermal states of light can be exploited to implement ghost imaging of simple objects. Based on those results, here we want to focus on the figures of merit that quantify the quality of the obtained images. More precisely, to compare the results of GI and DGI, we consider C and SNR, defined as
 \cite{ferri10,losero,iskhakov,ragy}
        \begin{eqnarray}
           {\rm C} &=& \sqrt{G_{im}(I)-G_{bg}(I)}\\
            {\rm SNR} &=& \frac{G_{im}(I)-G_{bg}(I)}{\sigma[G_{bg}(I)]},
        \end{eqnarray}
where $G_{im}(I)$ is the portion of either GI or DGI correlation matrix reconstructing the image of the object, while $G_{bg}(I)$ is the portion corresponding to the background. To investigate these parameters as functions of the number of speckles illuminating the object, we decided to perform a numerical simulation. This procedure enabled us to test advantages and disadvantages of the considered light sources in a more ideal situation, that is avoiding the experimental limitations related to noise (both from electronics and spurious light), imperfect single-mode selection, and non-optimal realization of the nonlinear interaction (realization of phase-matching condition and conversion efficiency).
 The simulation of both cases A and B was performed using LabVIEW. The program produced a speckle pattern by utilizing delta-correlated random matrices, subsequently convolved with a Gaussian distribution, resulting in a Gaussian field \cite{goodman}. 
This procedure was repeated many times in order to produce $10^5$ images. To generate the superthermal light source of case A, we selected a portion of the field by means of a virtual pin-hole and used 100 particles, randomly moving in the three-dimensional space, as scattering centers to diffuse the light exiting the pin-hole. A virtual CCD camera (200 x 200 pixels, 2 mm x 2 mm size) was used to monitor the obtained speckled-speckle field.
To produce the superthermal light source of case B, the matrices corresponding to the $10^5$ realizations of the speckle field were squared pixel by pixel to simulate a perfect second-harmonic generation. The same virtual camera as that in case A was used to detect the obtained second-harmonic field. 
In Fig.~\ref{fig:confrontoSNRandC} we show the values of C (see panel (a)) and SNR (see panel (b)) as functions of the number of speckles illuminating the object and belonging to superthermal light in case A (red symbols) and superthermal light in case B (black symbols). For the sake of comparison, in the same panels we show the values of C and SNR for thermal light (cyan symbols). First of all, we notice that for all the light sources the values of the figures of merit obtained with DGI and GI are identical, except for the SNR in case A. In that case, we have already demonstrated that DGI is better than GI \cite{cassina23}, since the former technique removes the contribution of the background, which is correlated (see the minimum value of the autocorrelation function in Fig.~\ref{fig:confrontoEXPsuper}). 
Secondly and more important, we highlight that, while the values of C of superthermal light are larger than that of thermal light, the SNR exhibits a more complex behavior. This is due to the fact that the contrast is, by definition, the square root of the difference between the maximum and minimum values of $g^2$, which in the case of a single speckle is equal to 1 for thermal light, $\sqrt{2}$ in case A and $\sqrt{5}$ in case B. On the contrary, the SNR, though being proportional to the difference between the maximum and minimum value of $g^2$, is also normalized to the standard deviation of the fluctuations of the background. This is the reason why the light source with a correlated background is endowed with the lowest values of SNR, while the other two cases, having the same kind of background, exhibit identical values.\\
To better visualize these results for C and SNR, in panels (b)-(d) of Figs.~\ref{fig:DGIsimulfixed} and \ref{fig:DGIsimulnorm} we present the DGI images of the binary mask of a llama shown in panels (a) of the same figures. All these images were obtained from the numerical simulations.
For both figures, the image in panel (b) was obtained with thermal light, the one in panel (c) with superthermal light in case A, and the one in panel (d) with superthermal light in case B. While in Fig.~\ref{fig:DGIsimulfixed} the images are shown with the same dynamical range, in Fig.~\ref{fig:DGIsimulnorm} each one is presented normalized to its maximum. 
The first choice (see Fig.~\ref{fig:DGIsimulfixed}) allows us to appreciate the different values of contrast of the images. In particular, the image obtained with thermal light has the lowest contrast, while the one obtained with superthermal light in case B the highest one. The second choice (see Fig.~\ref{fig:DGIsimulnorm}) emphasizes the different values of SNR. In fact, it is evident that the image obtained with superthermal light in case A is much noisier than the other two, which are very similar to each other.\\
All these results lead us to conclude that for practical applications, in which the SNR represents the most crucial parameter to certify the good quality of an image \cite{brida,losero,samantaray}, the use of thermal light represents the best choice since it is easier to produce with respect to superthermal light in case B. Possibly, the use of second-harmonic generation in the GI context could help if the medium used to diffuse light may be damaged by the wavelength corresponding to second harmonics \cite{lemos,allevi22}. Moreover, the light source in case B could be preferred to thermal light in application fields in which the SNR is not sufficient to discriminate the good quality of the reconstructed image, but also C must be taken into account. On the contrary, it seems that the light source in case A is less useful: its values of SNR are lower than those of the other two cases, and its values of C are lower than the corresponding ones in case B. Finally, we notice that both C and SNR are decreasing functions of the number of speckles used to illuminate the object. In particular, the colored lines shown in Fig.~\ref{fig:confrontoSNRandC} represent the fitting functions according to a power law \cite{brida}.\\
In conclusions, in this Letter we have investigated the possibility to perform GI and DGI schemes by employing two superthermal light sources obtained applying either linear manipulations or nonlinear interactions to a thermal field. Since the main resource of these techniques is based on the existence of intensity correlations, the larger statistical fluctuations of these light sources compared to thermal light seem to suggest that they could improve the quality of GI and DGI images. This advantage can be quantitatively proved by calculating the contrast C of the image, which attains values larger than the ones obtained with thermal light. On the contrary, the calculation of SNR, which is a more critical parameter with respect to C for real applications \cite{brida,losero,samantaray}, demonstrates that there is no advantage in using superthermal instead of thermal light. In fact, the results obtained for superthermal light from second harmonics are identical to those achieved with thermal light. Moreover, the implementation of superthermal light is experimentally more demanding and complex than the generation of a standard thermal light field. Hence, only in specific situations, where there is a practical need to convert light from one wavelength to another one, such as for diffusers that could be damaged by some specific wavelengths, the use of second-harmonic process might be preferred.
\begin{center}
\textbf{Declaration of competing interest}    
\end{center}
The authors declare that they have no known competing financial interests or personal relationships that could have appeared to influence the work reported in this paper.
\begin{center}
\textbf{DCRediT authorship contribution statement}    
\end{center}
S.C.: Investigation, Data curation, Methodology, Writing – original draft, Writing review $\&$ editing; G.C.: Software, Data curation, Methodology, Writing – original draft, Writing review $\&$ editing; M.L.: Conceptualization, Validation, Writing – original draft, Writing review $\&$ editing; M.B.: Conceptualization, Validation, Writing original draft, Writing review $\&$ editing; A.A.: Conceptualization,  Methodology, Validation, Writing – original draft, Writing review $\&$ editing.





\end{document}